%% file: uvz_treyer_etal98.tex
\documentstyle[psfig]{mn}

\journal{\Large \it Version: 28 May 1998}

\title[An Ultraviolet-Selected Galaxy Redshift Survey]
{An Ultraviolet-Selected Galaxy Redshift Survey: New Estimates of
the Local Star Formation Rate}
\author[M. Treyer et al.]{
{\LARGE \rm Marie A. Treyer$^{1,2}$, Richard S. Ellis$^2$, Bruno Milliard$^3$,
Jos\'e Donas$^3$, Terry J. Bridges$^4$} \\
$^1$ Astrophysikalisches Institut Potsdam, An der Sternwarte 16, 14482 Potsdam, Germany \\
$^2$ Institute of Astronomy, Madingley Road, Cambridge CB3 OHA, UK \\
$^3$ Laboratoire d'Astronomie Spatiale, Traverse du Siphon, 13376 Marseille, France \\
$^4$ Royal Greenwich Observatory, Madingley Road, Cambridge CB3 0EZ, UK 
}

\date {Accepted ---. Received ----; in original form ----.}

\newcommand{\beq}{\begin{equation}}
\newcommand{\eeq}{\end{equation}}
\def\lsim{\mathrel{\rlap{\lower 3pt\hbox{$\mathchar"218$}}
     \raise 2.0pt\hbox{$\mathchar"13C$}}}
\def\gsim{\mathrel{\rlap{\lower 3pt\hbox{$\mathchar"218$}}
     \raise 2.0pt\hbox{$\mathchar"13E$}}}

\def \d{{\rm d}}

\begin{document}
\textheight=8.5truein
\label{firstpage}
\maketitle

\begin{abstract}
We present the first results of an ongoing spectroscopic survey of
galaxies selected in the rest-frame ultraviolet (UV). The source
catalogue has been constructed from a flux-limited sample of stars,
galaxies and QSOs imaged at 2000 \AA\ in Selected Area 57 with the FOCA
balloon-borne imaging camera (Milliard et al.~1992). Accurate positions
for the UV sources have been obtained by matching with optical
counterparts using APM scans of the Palomar Sky Survey limited at
$B$=20.5. Here we present results derived from optical spectroscopy
conducted with the WIYN and WHT telescopes for 142 faint sources. The
redshift distribution for this UV-selected sample extends over
$0<z<0.5$ and a high fraction of the sources show intense nebular
emission lines and UV-optical colours bluer than normal Hubble sequence
galaxies. Such UV-selected surveys are thus a very efficient way to
locate and study intermediate redshift galaxies undergoing intense star
formation. Although our sample is currently small, we derive a
rest-frame UV luminosity function with a steep faint end slope
consistent with that found for late-type galaxies in optical samples.
However, the integrated luminosity density derived implies a
volume-averaged star formation rate higher than other recent estimates,
assuming a normal initial mass function. If representative of other UV
fields, as suggested by UV number count studies, our data implies the
local abundance of star-forming galaxies may have been underestimated
and consequently claims for strong evolution in the global star
formation rate in the range 0$<z<$1 overstated. An intensive study of a
large UV-selected sample is likely to reveal important information on
the declining population of star-forming galaxies of all types.
\end{abstract}


\section{Introduction}

The ultraviolet (UV) spectral region remains one of the last
astronomical windows to be systematically explored, particularly for
normal galaxies at low and intermediate redshift. Since the
pioneering work of Pence (1976) and Coleman, Wu and Weedman (1980),
good progress has been made in deriving the UV spectral energy
distribution of bright galaxies via the International Ultraviolet
Explorer (IUE, Ellis et al.~1982, King \& Ellis 1985, Burstein et al.~1988,
Kinney et al.~1993, 1996; Schmitt et al.~1997) and the Hopkins
Ultraviolet Telescope (HUT, Brown et al.~1997). However, in these studies,
only optically-selected sources have been chosen for UV follow-up. There
remains no systematic census of the local universe conducted in
the UV. Paradoxically, more is known about the UV properties of distant
galaxies through the study of very faint, redshifted sources. Indeed,
in interpreting distant sources, considerable reliance is placed on 
our scant knowledge of the UV properties of nearby galaxies.

The need for a comprehensive survey of the UV universe has become more acute 
as a result of recent progress made towards constraining the star formation
history of normal field galaxies (see Madau et al.~1996, Madau 1997). The rapid
increase in the rest-frame blue luminosity density to redshifts $z \simeq$1
(Lilly et al.~1995, Ellis et al.~1996) contrasts with the modest density of
star-forming systems selected using Lyman-limit techniques at $z>$2.3 
(Steidel et al.~1996, Cowie \& Hu 1998). It is tempting to connect these 
results and infer a low mean redshift of star formation consistent with 
QSO absorber estimates of gas depletion and chemical enrichment rates 
(Fall et al.~1996) and theories based on dark matter halos which merge
hierarchically (Kauffmann et al.~1994, Cole et al 1995, Baugh et al.~1997).

Although the latter result seems appealing, it remains unclear whether
the various star formation rates (SFR) that have been used to construct
this empirical result are being inferred consistently. The different 
diagnostics include nebular emission lines (Gallego et al.~1995, 
Tresse \& Maddox 1997) and the near UV continuum flux (Lilly et al.~1995, 
Ellis et al.~1996, Steidel et al.~1996). In most cases, these estimates 
are based on only the bright portion of the relevant luminosity function 
(LF), or on samples short of those ideally suited to the task.  Moreover, 
the vexing question of dust extinction remains unclear, particularly at 
high redshift where there is conflicting evidence on the corrections 
necessary (Pettini et al 1997, Meurer et al.~1997). Preliminary observations
by the ISO satellite (Rowan-Robinson et al.~1997) and by sub-mm detectors 
(Smail et al.~1997) suggest significant corrections may also be necessary 
to account for obscured sources.

The existing and forthcoming data from various sources could be better
interpreted if there was a greater overlap between the low/moderate and
high $z$ SF diagnostics. Of particular importance is the very dramatic
evolution in the observed luminosity density, ${\cal L}$, for the redshift
interval 0$<z<$1 --- ${\cal L}_{2800} \propto (1+z)^4$ --- which corresponds 
to an order of magnitude decline in the volume-average SFR (Lilly et al
1996). To understand the physical significance of such a rapid fading
in the UV light of the Universe requires independent and more sensitive 
data at moderate redshift. 

This paper presents the first results from an ongoing deep UV-selected
galaxy redshift survey designed to span the redshift range
0$<z<$0.5.  The overall aim is to provide an independent measure of the
declining SFR and a robust estimate of the present star formation
density. The selection of galaxies in the far UV (at 2000 \AA) leads to
a much greater sensitivity to the instantaneous star formation than is
the case with optically-selected samples. Moreover, the greater sensitivity
to extinction at short wavelengths is advantagous in constraining the
possible effects of dust in the star-forming population.

Progress on these questions has been hindered by the absence of any 
wide-field facility with which to conduct deep UV field surveys. Neither 
IUE, HUT nor the Hubble Space Telescope (HST) can be efficiently used in 
a survey mode. However, using the FOCA balloon-borne camera, Milliard et
al.~(1992) have constructed the first deep galaxy survey at 2000 \AA. 
The availability of the FOCA dataset has encouraged us to conduct a 
systematic UV-selected galaxy redshift survey to address the above 
questions. We begin here with a concerted attack on the field Selected
Area 57 which contains some of the deepest and most complete FOCA 
photomotric data.

A plan of the paper follows. In $\S$2 we discuss the construction of 
the photometric and astrometric database and the spectroscopic exposures 
and associated data reduction which lead to a redshift catalogue. In $\S$3 
we discuss the UV-optical colour distributions and compare these with 
expectations based on the optical $k$-corrections in current use. In $\S$4 we 
construct the rest-frame UV luminosity function and use it to provide 
an estimate of the volume averaged star-formation rate over the redshift
interval 0$<z<$0.5. In $\S$5, we summarise our principal conclusions.

\section{The Data}

Our redshift survey is based on images taken and analysed by the FOCA
balloon-borne wide-field UV camera. Further details of the instrument
can be found in Milliard et al.~(1992). Briefly, the experiment consists
of a 40-cm Cassegrain telescope equipped with an image intensifier
coupled to a IIaO photographic emulsion. The telescope is flown on a
stratospheric gondola stabilised to within a radius of 2 arcsec rms. The 
impressive feature of the experiment is the field of view (2.3$^{\circ}$ 
in its f/2.56 wide-field mode) which, together with the depth attained in
$\simeq$3000 sec exposures, makes it ideal for this programme. Photometric
selection via the FOCA 2000 filter whose spectral response approximates
a Gaussian centred at 2015 \AA\ with a FWHM of 188 \AA\ . The 2000
\AA\ galaxy count slope based on data taken in 4 fields is surprisingly
steep suggesting either recent evolution for some fraction of the population
(Armand \& Milliard 1994) or a significant error in the mean
$k$-correction in current use. One of the most intriguing features of
this unique dataset is the high fraction of UV-selected galaxies that
are much bluer in UV-optical than can be accounted via normal Hubble sequence
galaxies (Donas et al.~1995). 

The most suitable field for study from Milliard et al's series of 4 exposures
was considered to be Selected Area 57 (SA57). This field, centred at 
RA=13 03 52.61 Dec=+29 20 30.07 (1950.0), has been studied with both the 
FOCA 1000 (f/2.56, 2.3$^{\circ}$) and FOCA 1500 (f/3.85, 1.55$^{\circ}$) modes
ensuring the deepest, most reliable and widest-field (2.3$^{\circ}$) 
catalogue. The limiting magnitude is $m_{UV}$=18.5 which corresponds to 
$B\simeq 20-21.5$ for late-type galaxies. The present results are derived 
from multi-fibre spectroscopic exposures for SA57 taken in March 1996 with 
the Hydra instrument on the 3.5m WIYN telescope\footnote{The WIYN Observatory 
is a joint facility of the University of Wisconsin-Madison, Indiana 
University, Yale University, and the National Optical Astronomy 
Observatories.} and, later, in April 1997 with the WYFFOS facility on the 
4.2m William Herschel Telescope.

Both the WIYN and WHT instruments offer $\simeq$100 fibres over a 1 degree
diameter field, and since the surface density of galaxies at $m_{UV}$=18.5 is
$\simeq$ 200 deg$^{-2}$, there is no shortage of sources for study. However,
the imaging resolution of the UV data is only 10/20 arcsec FWHM depending,
respectively, on whether the FOCA 1500/1000 catalogues are used. The
astrometric precision is therefore typically 3-4.5 arcsec rms and insufficient
for defining a spectroscopic target list. Furthermore, the UV data cannot, on
its own, be used to discriminate between stars and galaxies. Accordingly, 
to overcome these limitations, we have matched the FOCA catalogues for SA57 
with Automated Plate Measuring (APM) machine scans of the Palomar Observatory 
Sky Survey (POSS) 103a-O and 103a-E plates. 

After eliminating the periphery of the FOCA images where the source
detection is less reliable, the FOCA 1500 catalogue consists of 643
sources with $m_{UV}\le$18.5 whereas the FOCA 1000 covers a wider field
and contains 805 sources. All sources were matched with APM scans of
the Palomar Sky Survey limited at $B$=20.5 using a search radius of 10
arcsec when detected by FOCA 1500 and 20 arcsec when detected by FOCA
1000 only.

Locating the appropriate optical counterpart for each UV source proved
occasionally troublesome on two accounts. Firstly, some of the faintest
UV detections have no obvious optical counterpart. Only 6\% of the FOCA
1000 sources are so affected but the fraction rises to $\simeq$30\% for
the deeper FOCA 1500 list. Clearly some fraction of the faintest UV
detections are either spurious or their optical magnitudes lie
fainter than $B$=20.5. We will return to this point in $\S$3.
 
The second problem is that, in many cases, more than one optical 
identification lies within the UV error circle. It is not obvious which is 
the correct target, or whether the UV flux arises from a combination of 
more than one source. This is most problematic for the FOCA 1000 catalogue 
where a search radius of 20 arcsec is necessary. 60\% of the FOCA 1000 sources
have more than one possible APM counterpart. This fraction reduces to only 9\%
when using a 10 arcsec search radius. For the FOCA 1500 sources 13\% have two
APM counterparts and 1\% have three counterparts.

In cases of multiple identifications, we selected the optical source
closest to the UV position. For the region where the FOCA 1000 and FOCA
1500 images overlap we checked the validity of this approach by
matching common sources and obtaining improved positions. We found that
selecting the optical ID closest to the FOCA 1000 position was
generally quite reliable. In practice the main confusion arising from
this problem is therefore the possibility of contamination of the UV
signal from more than one source.

Ideally, it would be desirable to spectroscopically study all sources,
regardless of whether they appear to be stars or galaxies. For example,
there might be a significant population of compact star-forming galaxy
lost by applying standard optical star/galaxy separation criteria.  In
practice, however, for the exploratory survey discussed here, this
approach was not considered practical because of the high fraction
($\simeq$30\%) of likely stars. Star/galaxy classification was thus
attempted both using the APM algorithms and via visual inspection. The
final image classifications were merged and disputed cases included as
potential galaxies. It should be noted that the possibility of further
extragalactic UV sources would only strengthen the principal conclusion
of this paper.

The FOCA photometry is discussed in detail by Milliard et al.~(1992)
and is defined according to a scheme similar, but not exactly equivalent,
to the ST system. Following Donas et al (1995), 
$m_{\lambda}$= - 2.5 log ($F_{\lambda}$ - 21.175, where $F_{\lambda}$ is in 
cgs units. The UV zero point is based on earlier stellar photometry in the 
range 10$<m_{UV}<$13 conducted using the SCAP 2000 experiment (Donas et 
al.~1987) and is accurate to better than 0.2 mag. Given the low UV background,
the principal uncertainty in the relative galaxy photometry arises from 
possible non-linearities in the photometric system and this is unlikely
to dominate the photon statistics. It is estimated that total photometric 
error of the UV data may reach $\pm$0.3-0.5 close to the limit of our
survey (Donas et al.~1995). The detected UV background is sufficiently 
faint that the FOCA magnitudes represent isophotal values at a surface
brightness of $\mu_{UV}$=27 arcsec$^{-2}$, i.e. sufficiently close to 
total for the sources of interest in this paper. 

The optical $B$ and $R$ photographic photometry was taken directly from the
POSS database and includes corrections for photographic saturation and 
isophotal losses (McMahon, personal communication). Given the poor angular 
resolution of the FOCA data, the UV-optical colours may be subject to a 
possible systematic offset compared to their idealised (aperture) 
equivalents. We discuss this question further in $\S$3 when presenting the 
distribution of UV-optical colours.

To commence the survey, we were allocated queue-scheduled time with the
Hydra multi-fibre spectrograph on the 3.5m WYIN telescope. The target
list of 71 galaxies (drawn from the FOCA 1000 list with $m_{UV}<$18.5)
and 13 sky fibres was configured using the WHYDRA utility. The WIYN
spectroscopic exposures of SA57 were taken on February 28 and 29 1996
in service mode. The multi-fibre spectrograph was used with the Simmons
Camera, the blue fibre cable and the Tek 2KC CCD. The 400-line grating
was used offering a wavelength range $\lambda\lambda$ 3500-6600
\AA\ and a spectral resolution of 7 \AA. On each night three 1800 s
field exposures were taken and these were bracketed with CuAr lamp
calibrations. The seeing was 1 arcsec throughout. The spectroscopic
data was processed within IRAF using the DOHYDRA package.  Dark frames
were subtracted and dome flatfielding performed using combined
exposures taken on the same night supplied with the service data. The
spectral data has not been flux calibrated.

In April 1997, we were allocated queue-scheduled time with the WYFFOS
multi-fibre spectrograph at the prime focus of the 4.2m William Herschel 
Telescope. In this case, the target list was configured for the Autofib2 fibre
positioner using the ${\tt af2\_configure}$ program. Targets were
selected from an improved FOCA list determined from the merger
of the FOCA 1000 and FOCA 1500 catalogues (as described above) after 
eliminating those with successful WIYN redshifts. In this case we used a 
300-line grating with the Tek6 1K CCD giving a wavelength range of 
$\lambda\lambda$3500-9000 \AA\ and a spectral resolution of 10 \AA. 
The greater wavelength coverage enabled the location of both [O II] 
and H$\alpha$ emission in many of the spectra. A 4$\times$1800s 
spectroscopic exposure was taken and a second exposure was begun with 
an independent Autofib2 configuration but the latter sequence was 
curtailed after one hour due to poor weather conditions. The WHT 
spectroscopic data was processed within IRAF using the package
purposely written for WYFFOS. Figure 1 presents a selection of the WHT
and WYIN spectra.

\begin{figure*}
\psfig{figure=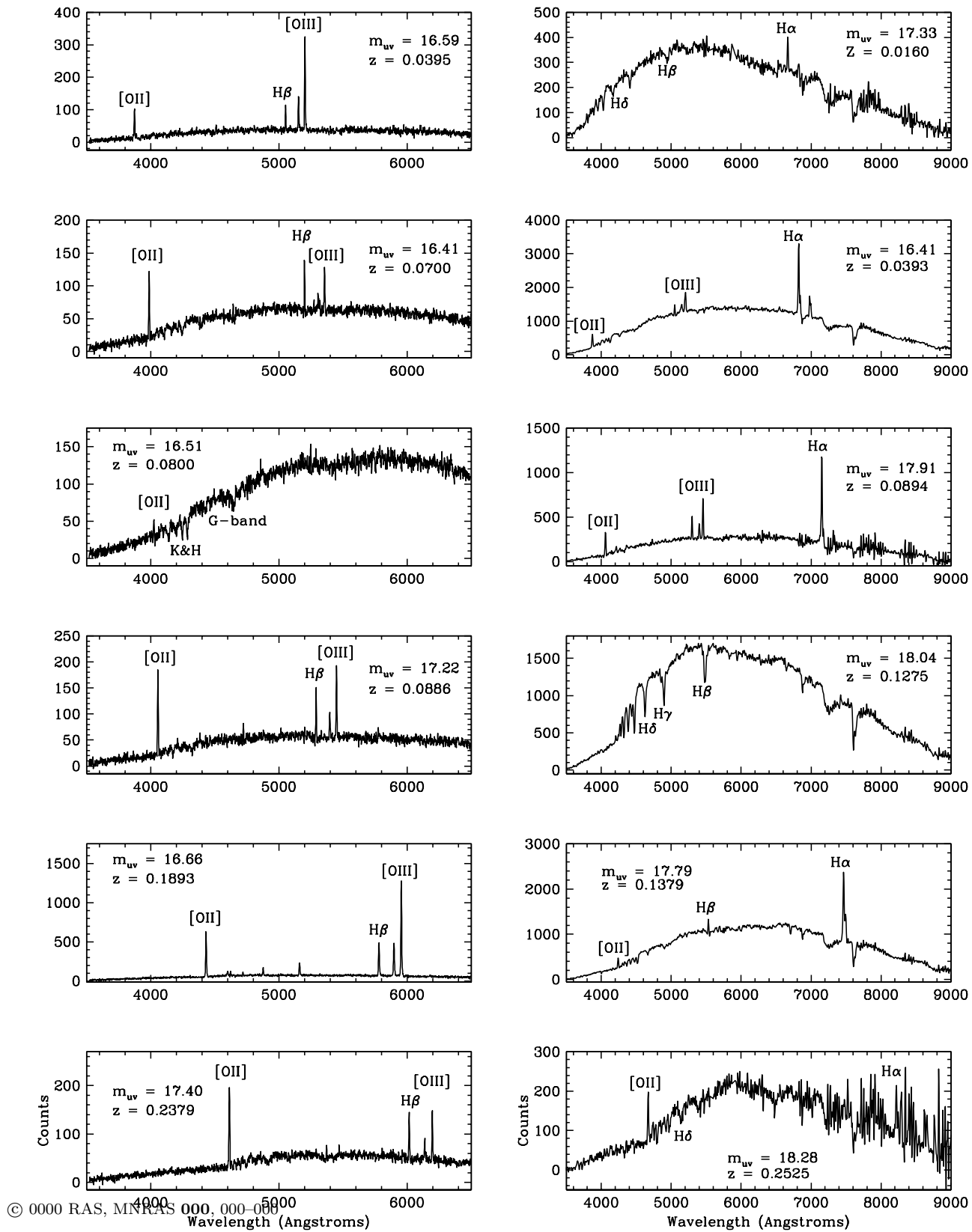,width=175mm}
\caption{A selection of optical spectra for galaxies in the UV survey.
The left panel refers to UV targets observed with the WIYN+Hydra 
combination. The right panel refers to those observed with the WHT+WYFFOS.}
\end{figure*}

Redshifts were measured by visual inspection and [O II] and H$\alpha$
equivalent widths determined using the ${\tt splot}$ facility within IRAF.
From the 3 exposures, reliable spectra were obtained for 142 sources
to $m_{UV}$=18.5. As expected, the incompleteness is unacceptably high
in the second, shorter, WHT exposure and so the bulk of the analysis
that follows has been conducted using the other two exposures yielding
a total of 128 spectra and an identification completeness of 90\%. Of
the 14 unidentified spectra in these two fields, 6 suffered from
technical difficulties in extraction unrelated to the signal/noise. The
formal incompleteness within these two fields is thus only 8/122 viz.
6\%. Within this complete sample of 122 sources, we found 103 galaxies,
5 broad-line AGN and 6 stars. A further 12 galaxies and 2 stars were 
identified from the second WHT exposure. 

The modest but non-zero contamination of our sample by AGN and stars
(8\% of the total sample) is a good indication that we have been
appropriately over-cautious in our star/galaxy classification
procedure. Without any star/galaxy separation the stellar fraction to
$m_{UV}$=18.5 would have been $\simeq$30\% (Milliard et al.~1992). At
first it seems perhaps surprising that that such a high fraction of the
compact contaminants are AGN. However, our spectroscopic AGN have
optical magnitudes and redshifts in the ranges 18.2$\le B \le$ 20.4
and  0.7$\le z \le$1.5 respectively and the expected number of UVX
QSO's in this magnitude range is $\simeq$23 deg$^2$ (Boyle, Shanks \&
Peterson 1988) i.e. $\simeq$10\% of the UV galaxy density to our
limit. 

For the remaining galaxies with redshifts, we can further restrict the sample
into those for which there is a reliable UV flux from both the FOCA 1000
and 1500 catalogues and an unambiguous optical ID with no confusion. 
Starting from a maximal sample of 115 galaxies across all 3 exposures (103 
if we restrict to the two well-exposed fields), 11 were retrospectively
determined to have unreliable UV fluxes and have been discarded from
the analysis. 19 galaxies belong to the double counterpart category and 2 are 
triple counterpart cases. The most secure analysis we can consider is 
therefore based on what will refer to as the {\it restricted sample} of 84 
galaxies with measured redshifts (75 in the two well-exposed fields). We 
will explore the effect of including the multiple counterpart cases in 
the discussion below. The larger sample of 104 galaxies (115 less 11
sources with poor UV fluxes) across all 3 fields will be referred to 
as the {\it full spectroscopic sample}.

A large fraction of the spectra reveal strong emission lines as expected if
the bulk of the UV-selected sources are star-forming galaxies. 46\% of
the galaxies have $W_{\lambda}$[OII]$>$15\AA, compared to $\sim 15\%$ in
optically-selected samples of comparable depth (Peterson et al.~1986,
Broadhurst, Ellis \& Shanks 1988, Heyl et al.~1997). Figure 2 indicates
the overall redshift distribution of the sample where the presence
of the Coma supercluster ($z$=0.023) can be seen. Table 1 presents
the final spectroscopic catalogue separately for the galaxies and
other sources. Multiple counterpart cases are identified via the column 
labelled `OC'. 

\begin{figure}
\psfig{figure=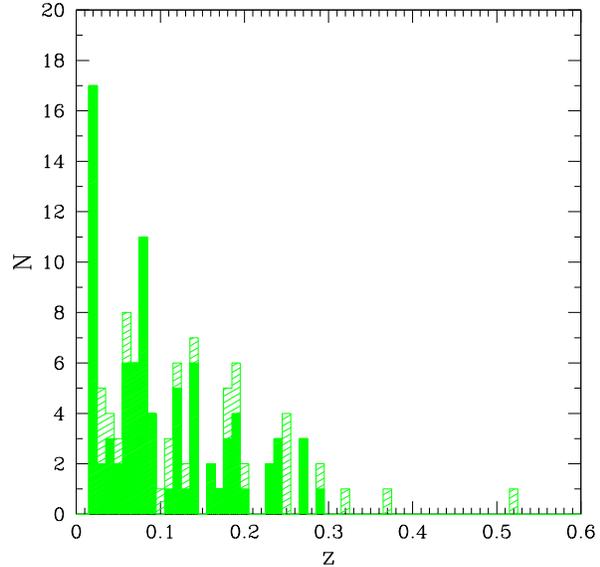,width=80mm}
\caption{The redshift distribution for the combined WIYN and WHT samples
limited at $m_{UV}$=18.5. The filled component refers to the restricted
sample and the shaded component represents those UV sources for which 
a choice of optical counterpart was available. The peak at $z\simeq$0.02
arises from the Coma supercluster.}
\end{figure}

\section{Optical-Ultraviolet Colour Distributions}

We now wish to explore the extent to which the UV-optical {\it colour 
distribution} observed for our UV-selected sample is consistent with that 
expected on the basis of models used in conventional analyses of the optical 
galaxy data (c.f. review by Ellis 1997). This is equivalent to addressing 
the question of whether the mean optical $k$-correction in current use
are appropriate. Although extensive spectral energy distributions 
(SEDs) are available for individual galaxies of known Hubble class 
from optically-selected samples (c.f. Kinney et al.~1996), we have little 
idea whether the available SEDs span the full range of the UV-emitting 
population and hence whether the UV-optical colour distribution is 
correctly inferred using optical data.

In fact, there have already been several indications that some galaxies 
exist with significantly stronger UV fluxes than would be expected from 
the normal range of optical SEDs. Armand \& Milliard (1994) found that 
the FOCA counts across several fields have a steeper $N(m)$ slope and 
higher normalisation than popular `no evolution' models would predict. 
This is surprising considering the modest redshift range probed in our 
survey. Moreover, Donas et al.~(1995) found a significant population of 
sources in their FOCA image of the Coma cluster with UV excesses of as 
much as 1-2 magnitudes compared to standard models that account for a 
range of optical $b-r$ colours. Fioc \& Rocca-Volmerange (1997) have 
postulated a connection between these UV-strong sources and the 
abundant population of emission line galaxies seen in deep redshift 
surveys. They suggest that sudden bursts of star formation superimposed 
on normally-evolving population may be needed to account for the 
anomalously strong UV flux.

The availability of redshifts for a complete sample of UV-selected
galaxies enables further progress to be made on these and related
questions. For comparison with our UV-optical colours we will refer to 
the model SEDs published by Poggianti (1997). These are particularly
well-suited to the question at hand because Poggianti selects her model
SEDs to match the observed broad-band colour data of galaxies of
various Hubble types (in contrast to Kinney et al.~(1996) who use
empirical aperture spectrophotometry). 

UV-optical colours have been predicted for the various Poggianti SEDs 
as a function of redshift using the appropriate filter functions for 
the FOCA 2000 \AA\ filter and for the Palomar photographic systems 
(hereafter $B$ and $R$). In comparing these predictions with our
UV-optical colours it is necessary to take account of the fact that
the UV photometry is defined on its own, non-standard system (see
$\S$2, Donas et al 1995) whereas our optical photometry is based
on the normal $\alpha$-Lyrae system. In order to maintain consistency
with earlier UV papers, we have chosen to reduce both our model
predictions and optical data to the same UV-based system. The correction
for the APM $B$-magnitudes (defined in the $\alpha$-Lyrae system) amounts
to 0.37 mag in the sense $b\equiv B_{corr}$=$B_{APM}$ - 0.37.

The UV$-b$ colour-redshift distribution for the full spectroscopic sample
is shown in Figure~3 together with, suitably corrected, predicted colours 
from the SEDs of Poggianti (1997). The two bluest predictions (labelled SB) 
are examples of starburst models kindly supplied by Dr Poggianti specifically 
for this comparison. Type SB1 (the redder case) assumes a burst of star 
formation of duration $10^8$ year immediately prior to the epoch 
of observation involving 30 \% of the galactic mass. Type SB2 (the bluer 
case) assumes a shorter $10^7$ year burst involving 80 \% of the galactic 
mass. The star formation history prior to these bursts is that of a passively
evolving system. These models provide a good illustration of the range
of UV-optical colours expected for normal initial mass functions.

\begin{figure}
\psfig{figure=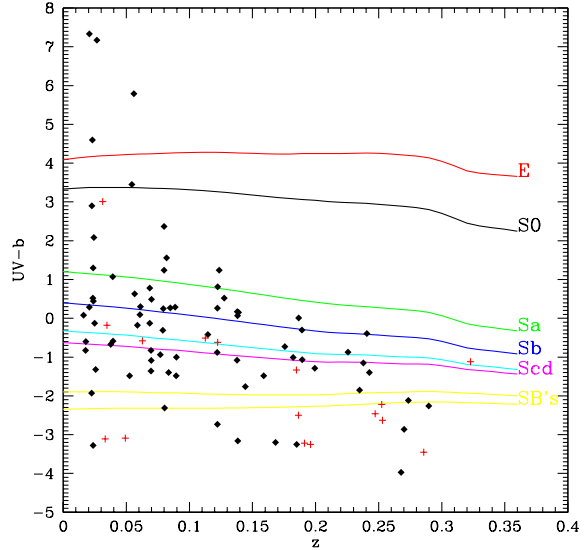,width=80mm}
\caption{The UV$-b$ redshift relation for the full spectroscopic 
sample. The curves refer to the predicted colour at each redshift
according to the model SEDs of Poggianti (1997) - see text for further
details, including the definition of our photometric system. Crosses 
represent the multiple counterpart cases.}
\end{figure}

The most obvious, and puzzling, feature of the distribution is that 
a significant fraction  ($\sim 10\%$) of UV sources have colours 
up to 2 magnitudes bluer than even the strongest starburst SED. This 
phenomenon is not associated solely with faint sources or those which 
suffer confusion in their optical identifications. The discrepancy is 
more striking when it is realised that the model SEDs are unreddened. 
Comparisons with other SEDs, such as the empirical set produced by
Kinney et al.~(1996) produce larger discrepancies. 

Before interpreting this figure it is appropriate to return to the question 
of a possible systematic offset between the UV and optical photometric 
systems. Such an offset would work in the sense of generating galaxies
with too much UV flux if, for example, the UV system referred to a 
deeper isophote than is the case in the optical. Whilst this is undoubtedly
a possibility, it is difficult to imagine effects of greater than 
0.2-0.3 mag (Metcalfe et al.~1995). Similarly, a zero point offset or
scale error in either photometric system would yield an offset, again
typically of no more than 0.2 mag (c.f. $\S$2) and, in the case of
scale errors, anomalous sources would primarily be at the faint end of
the apparent magnitude range in the offending system. The trends we seek 
to explain have signatures that are much greater than the possible 0.2-0.3 
mag uncertainties.
 
Concerning the accuracy of the optical photometry, 
it is conceivable that some of the anomalously UV bright
sources represent cases where the POSS magnitudes are poorly determined.
We consider this to be unlikely given that, to reconcile the colours
with normal SEDS, the true optical magnitudes for these sources would 
have to be $B\simeq$19-20 where the reliability of the POSS photometry
is usually fairly good.

Importantly, however, the presence of {\it some} UV strong sources with 
UV-$B\simeq -3$ implies that our optical magnitude limit of $B$=20.5 may
be inadequately shallow and that this intriguing population of UV-strong
sources is {\it underestimated} in our survey. This is relevant in further
consideration of the 30\% of FOCA 1500 catalogue limited at $m_{UV}$=18.5, 
which has no optical counterpart. Deeper imaging is required to understand
whether a significant proportion of these are genuine sources. Of course
this becomes increasingly difficult to use as the surface density of
optical sources (and the likelihood of false associations) increases.

Regardless of the true number of sources with anomalous colours to 
$m_{UV}$=18.5, we seek a physical explanation for these extreme 
objects. Three explanations occur to us. Firstly, the model SEDs 
produced by Poggianti refer to calculations based on time steps of 
10$^7$ years or greater using normal mass function and solar metallicity. 
With a similar IMF and metallicity, a 1 Myr old starburst can be somewhat
bluer (Leitherer \& Heckman 1995). Stronger UV continua can also be produced 
for a limited duration in intense short-lived bursts with lower metallicity 
(Charlot 1996). The abundance of UV strong sources (which may be 
underestimated in the present sample) would then imply a larger 
population of sources undergoing such activity. Explanations based on 
abnormal IMFs are obviously possible though impossible to verify without 
UV spectra.
                                            
Secondly, conceivably some fraction of the UV light of these galaxies arises
from a non-thermal source. Tresse et al.~(1996) has argued that between
8-17\% of a sample of $z<$0.3 $I<$22 galaxies have emission line ratios
consistent with Seyfert2 or LINERs. Although there is no indication of 
broad emission lines in those UV sources in question, conceivably this 
is masked by a strong star-forming component as well as by the overall 
poor signal/noise in the exposures which precludes precise line
ratio measurements. Again, only UV spectra could clarify this possibility.

Alternatively, it is conceivable that the UV emission is somehow
independent of the optical radiation. A mismatch of apertures seems
unlikely for such a high fraction of our sample. It is interesting
to note, however, that the correlation between the UV-$b$ colour and 
equivalent widths of the nebular emission lines is not that strong (Figure 4). 
Although the galaxies with the most intense $H\alpha$ emission lines are 
generally stronger in the UV, there is a large amount of scatter. This is 
probably due to the fact that the equivalent widths are poorly determined
in all but the best spectra. 

\begin{figure}
\psfig{figure=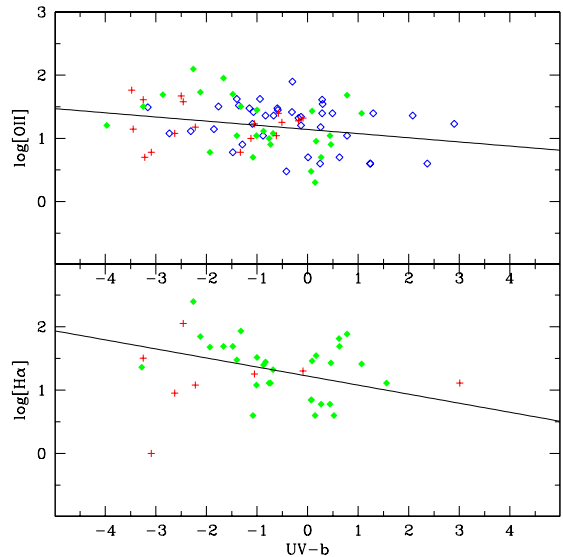,width=80mm}
\caption{The correlation of the observed UV-$b$ colour with 
the equivalent  width
of the nebular emission lines for the full spectroscopic sample: (a) [O II]
3727, (b) H$\alpha$.  Open symbols refer to the WIYN data, filed symbols to the
WHT data.  Crosses are the multiple counterpart cases.}
\end{figure}

In summary, no single convincing physical explanation can yet be found for 
these UV-strong sources and, perhaps most importantly, their numbers
may be underestimated in the current sample in view of the
magnitude limit of the optical data. 

The effect of this population on the UV luminosity function (LF) will be 
discussed in detail in $\S$5 but the mismatch between our expectations
based on optically-selected samples and the current results can also be 
judged by comparing the relative UV-$b$ colour distribution for our 
$m_{UV}<$18.5 sample with that expected on the basis of the 
King \& Ellis (1985) optical LFs differentiated by spectral class and 
coupled to $k$-corrections based on the Poggianti SEDs (Figure 5). 

Such model predictions are the norm for optical work (Ellis 1997) and
can provide good matches to the {\it optical} colour distributions of 
moderately faint samples (Metcalfe et al.~1991). However, as Figure 5 shows, 
these optically-based predictions cannot account for the full extent of the 
UV-$b$ colour distribution. The model also underestimates the observed 
number of UV galaxies brighter than 18.5 by a factor which depends
on the absolute normalisation of the optical LF (c.f. Ellis 1997) but
which is at least $\simeq$2-3. 

It is also interesting to consider whether this discrepancy can be resolved 
by adopting a steeper faint end slope for the optical luminosity function 
for late type star forming galaxies as proposed by Marzke \& Da Costa (1997) 
on the basis of their Southern Sky Redshift Survey. Adopting a faint end 
Schechter slope of $\alpha$=-1.5 for this subset of the population 
(c.f. -1.0 for the King \& Ellis LF), {\it and} arbitrarily giving all
of these sources the UV-$b$ colours of starburst galaxies, it is possible
to boost the number density of UV-selected objects by a considerable
factor close to the observed level. However, as Figure 5 shows, the colour 
distribution would then be considerably skewed to blue sources. Even allowing
for the possibility that the bluest sources are under-represented, this
prediction is incompatible with the observed colours. Regardless of the
final solution to this dilemma, it is clear that the optically-based
data cannot easily reproduce the UV colour distribution and counts.

\begin{figure}
\psfig{figure=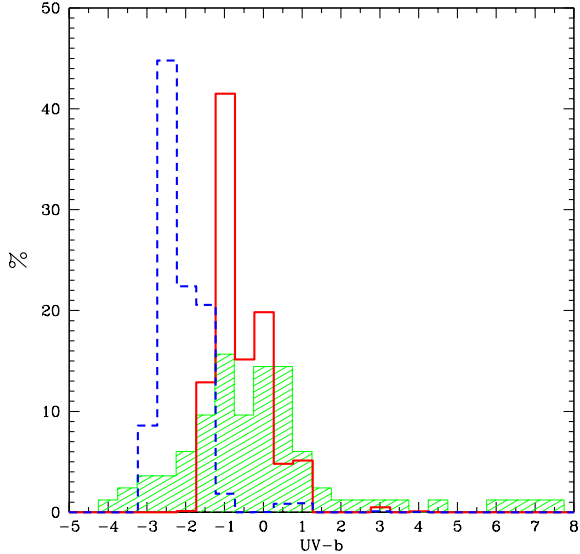,width=80mm}
\caption{The UV$-b$ colour distribution for the full spectroscopic
sample (shaded histogram) compared with model predictions based on 
the optical luminosity 
functions of King \& Ellis (1995) and the spectral energy distributions of 
Poggianti (1997) (solid line). This model fails to reproduce the bluest
sources and underestimates the UV number counts by a significant factor. 
The dashed line adopts a steeper faint end slope for the luminosity function 
and bluer UV-$b$ colours for the late type galaxies. Such a model can 
reproduce the UV counts but at the expense of a significant shift in mean 
UV-$b$ colour.}
\end{figure}

One advantage of the slow change in predicted UV-$b$ colour with redshift 
(Figure~3) is that it allows us to reliably use the observed colour to 
infer a spectral class for each galaxy essential for inferring a UV 
$k$-correction necessary for deriving the rest-frame UV luminosity function 
($\S$4). Admittedly, this assumes that reddening is not significant in
the interpretation of our colours, a point we return to in $\S$4.
In practical terms, we allocated a spectral class according to 
the scheme (E/S0, Sa, Sb, Scd, SB) using Poggianti's SEDs. The absolute 
magnitude then follows using:
\beq
M_{UV}=m_{UV} - 5{\rm log} d_L(z)-25 - k_{UV}(z),
\eeq
where $d_L(z)$ is the luminosity distance at redshift $z$. We assume 
$\Omega=1$ and $H_0=100~h~{\rm km~s^{-1}Mpc^{-1}}$ in all calculations. 

The UV $k$-corrections are quite small ($|k_{UV}(z)|<0.4$) and almost 
type-independent for the spiral classes which dominate the sample. This
is because of the narrowness of the FOCA 2000 \AA\ filter response function 
and the slowly-changing SEDs in both wavelength regions for galaxies
with weak Balmer discontinuities.

\section{The UV Luminosity Function and Local Star Formation Density}

We now turn to the determination of the rest-frame UV luminosity function (LF).
Before doing so we need to issue some important caveats. Firstly, clearly with 
only a single selected area, the current results must be regarded as 
preliminary. Secondly, it must be remembered that the LF is strictly that
derived for a selection criterion which includes an optical limit of $B<$20.5.
A significant proportion of sources with $m_{UV}<$18.5 have no optical IDs;
these will generally be the bluest sources and thus the LF derived may not be
truly representative of that for a strict UV-limited sample even when
normalised to the total  counts. Similarly by eliminating the multiple
counterparts we may likewise be biasing the LF against certain classes of
sources such as interacting galaxies. We can investigate some of these biases
but it is not yet possible to correct for them properly in the analysis.

We adopt the traditional 
$V_{max}$ method for the LF derivation (e.g. Felten 1977). The number 
density of galaxies with magnitude $M$ is given by:
\beq
\phi(M) \d M \propto \sum {\d M \over V_{max}(M,i) },
\eeq
where the sum is extended over all galaxies with magnitude 
$M\pm \d M/2$, and $V_{max}(M)$ is the maximum volume in which type 
$i$ galaxies with absolute magnitude $M$ are observable, i.e. satisfying
$m(M)\le18.5$ in Eq.~1. As is customary for small samples,
the normalisation of the LF is adjusted to fit the observed number 
counts (Milliard et al.~1992) assuming the restricted sample is 
representative of the entire source distribution. Since the mean counts
represent the average of FOCA exposures for several fields, this enables 
us to minimise uncertainties in normalisation which would otherwise
be based on a single field.

As our sample extends over a significant redshift range, prior to
deriving a single LF, it is worth considering whether there is an 
evolutionary trend. Such a trend might be expected from the results
of Ellis et al.~(1996) where a significant change in the star forming
component of the {\it optical} LF was detected. An important advantage 
of the UV selection over that in the optical (and apparent in Figure~3)
is that the apparent mixture of spectral classes is unlikely to be
significantly different at $z\simeq 0.3$ c.f. $z$=0. Accordingly, a 
simple $V/V_{max}$ distribution can be used to determine whether an unevolving 
LF is an appropriate representation of the data gathered so far (or
conversely, assuming no evolution, whether the sample is statistically
representative and complete). The resulting distribution of $V/V_{max}$ is 
shown in Figure 6 for the two substantially complete fields, after 
subtracting those sources lying in the redshift range of the Coma cluster 
($0.0225<z<0.0255$). Although the samples are small and the mean values
somewhat uncertain, there is clearly no evidence for strong evolution in 
the distribution. Rather, the mean values for both fields are slightly
less than 0.5. The overall mean is 0.42$\pm$0.03, i.e. non-uniform with
a marginal significance. This effect appears to arise primarily from
the WIYN field where the incompleteness is a bit worse and where the 
selection was based on the inferior FOCA 1000 photometry.

\begin{figure}
\psfig{figure=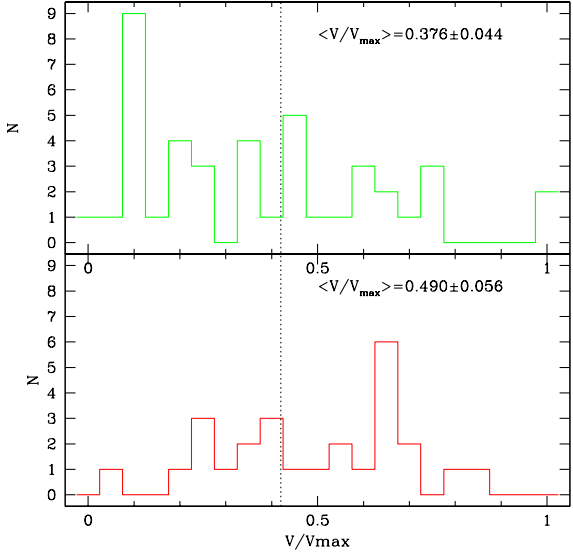,width=80mm}
\caption{The distribution of $V/V_{max}$ for the WIYN field (upper panel) 
and the WHT field (lower panel). Mean values are indicated for each sample.
The overall mean is shown as a dotted line.}
\end{figure}

The derived UV luminosity function (assuming $H_0=100$ kms s$^{-1}$
Mpc$^{-1}$) is presented in Figure 7 together with the best fit
Schechter (1976) function with the following parameters (errors
represent formal statistical uncertainties):

\beq
\begin{array}{rl}
\alpha &= -1.62~[+0.16,-0.21] \\
M_\star &= -20.47~[+0.31,-0.29] +5{\rm log}(h).
\end{array}
\eeq
i.e.
$L_\star= 6.16^{+1.88}_{-1.53} \times 10^{39}~ h^{-2}~{\rm ergs~s^{-1}\AA^{-1} }$.

The faint end slope of the UV LF is significantly steeper than that of 
the overall optical equivalent ($\alpha\simeq$-1.1 c.f. Ellis 1997) but 
comparable to some estimates of that for the star-forming late-type galaxies 
(Marzke et al.~1994). We tested the robustness of this result by excising 
low redshift sources with $z<$0.0255 in case additional large scale structure
might produce this effect. In this case the slope decreased to 1.52, i.e.
within the error bars of the above estimate. Adding the Coma members raises 
the slope to 1.76 (again within the error bars). 

Turning to the normalisation we obtain $\phi_\star = 9.12~\times 
10^{-3\pm0.24}~h^3~{\rm Mpc}^{-3}$ if we adopt the $m_{UV}<$18.5 surface
density of Milliard et al.~(1992). This is shown as the solid line 
in Figure~7. The dashed line indicates the model UV LF assuming all of 
the Marzke et al.~late type galaxies have UV-optical colours of SB galaxies.
Although an adequate representation of the data, as we saw in $\S$3, the 
colour distribution predicted by such a model is too blue.

\begin{figure}
\psfig{figure=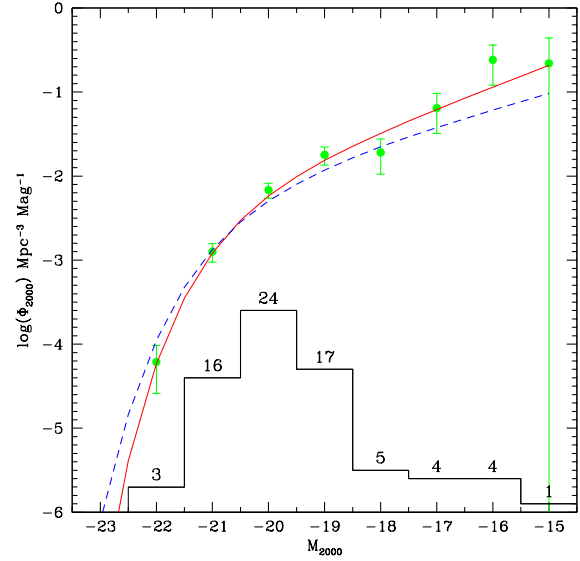,width=80mm}
\caption{The UV luminosity function derived from the present redshift 
survey. The solid line shows a best fit Schechter function derived
using a $V_{max}$ method normalised to the total Milliard et al.~(1992)
counts. The dashed line is a predicted LF based on allocating starburst 
UV-optical colours to Marzke et al's (1994) optical LF for late type 
galaxies. The histogram shows the number of galaxies contributing to each 
magnitude bin.}
\end{figure}

Addressing the question of the effect of an optical pre-selection
($B<$20.5) on the UV survey, Figure 8 compares the absolute magnitude
distribution of the restricted sample (i.e. UV sources having a unique
optical counterpart) with that for the bluest sources (UV-$b<-1$ 
corresponding to Scd and later) and that for the multiple counterpart 
objects in the full spectroscopic sample. As the relative distributions 
are similar, this would appear to indicate the main effect of omitting 
fainter optical sources and multiple counterparts is recovered in the 
final normalisation, i.e. the {\it shape} of the LF will be unaffected.

\begin{figure}
\psfig{figure=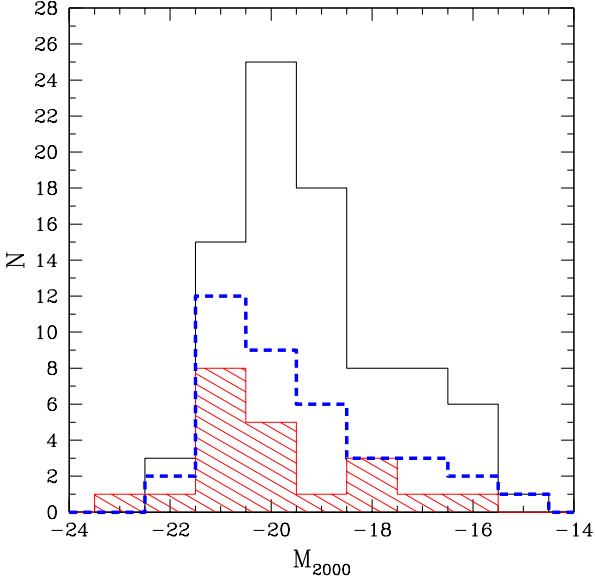,width=80mm}
\caption{The absolute magnitude distribution for the restricted sample,
i.e. the UV sources having a unique optical counterpart (solid line)
compared to that of populations drawn from the full spectroscopic
sample that may be inadequately represented in it.  The thick dashed
line represents the distribution of the bluest sources (UV-$b<-1$), and
the shaded histogram that of the multiple counterparts.
}
\end{figure}

Integrating the measured luminosity function over the observed 
range, $M_{UV}< -15$, yields a `resolved' luminosity density at 
2000 \AA\ of: 
\beq 
{\cal L}_{2000}^{res}= 1.53^{+0.21}_{-0.24} \times 10^{26}~h~{\rm ergs~s^{-1}Hz^{-1}Mpc^{-3}}
\eeq
Integrating the steep LF beyond the measured range
increases this estimate by $\sim$ 20 \%:
\beq
{\cal L}_{2000}^{tot}= 1.86^{+0.76}_{-0.45} \times 10^{26}~h~ {\rm ergs~s^{-1}Hz^{-1}Mpc^{-3}}
\eeq
The starburst population, defined again as galaxies with UV-$b<-1$ 
(cf. Figs.~4), contributes $\sim$ 40\% of the total light density.  
These estimates take into account the fact that some proportion of
the UV sources were not observed due to a lack of any optical ID
brighter than $B<$20.5 provided their distribution in luminosity
is no different from those present in the restricted sample (Fig. 8).

As dust may absorb a significant fraction of the UV photons within 
the galaxies, the observed UV luminosity density clearly represents 
a lower limit to that emitted. Whereas optically-selected galaxies 
are only moderately affected by dust, starburst galaxies such as those 
dominating the present sample may be more obscured. From a sample of 
local starburst galaxies for which UV to FIR photometric data are available, 
Buat \& Burgarella (1997) have derived a mean extinction of $\sim 1.2$ mag 
at 2000 \AA\ (although with a large dispersion). Likewise, Pettini et 
al.~(1998) estimate similar amounts of dust extinction in high redshift 
starbursts. Applied to each UV galaxy, such extinction would amount to 
increasing the observed luminosity density by a factor of over 2.5.
Other authors have argued for even greater absorption factors 
(e.g.~Meurer et al.~1997). However, for the purpose of comparison 
with other measurements as presented by Madau (1997), we adopt a 
correction factor of 1.8. This corresponds to the value found by 
interpolating the more modest corrections applied by Madau to the 
luminosity densities observed at 2800 and 1500 \AA\ respectively.

The UV continuum emission in all but the oldest galaxies is dominated 
by short-lived massive main sequence stars of spectral class 
late-O/early-B and, modulo the effects of dust, provides one of the
most direct measures of the instantaneous star-formation rate (SFR)
(Madau 1997). For a Salpeter IMF extending from 0.1 to 125 
${\rm M}_{\odot}$, and assuming solar metallicity, the SFR in units 
of ${\rm M_{\odot} yr^{-1}Mpc^{-3}}$, {\it uncorrected for dust 
extinction}, is: 
\beq
{\rm log}(SFR)=  {\rm log}({\cal L}_{2000}^{tot})-27.9=-1.62^{+0.15}_{-0.12}
\eeq
Assuming a Scalo IMF, which is significantly less rich in massive 
stars, increases this value by +0.3 dex (Madau 1997).

In the approximation of instantaneous recycling (Tinsley 1980),
the ejection rate of newly synthesized heavy elements 
($Z\ge6$) is directly proportional to the star formation rate.
The conversion factor is very sensitive to the slope 
and lower-mass cutoff of the IMF. With our present assumption
of a Salpeter IMF, a $0.1-125$ ${\rm M_{\odot}}$ mass range and 
initial solar metallicity, our dust-corrected estimate yields: 
\beq 
\dot \rho_Z = 0.024\times SFR=
1.03 ^{+0.43}_{-0.25}\times 10^{-3} {\rm ~M_{\odot} yr^{-1}Mpc^{-3}}.
\eeq

Figure 9 shows recent estimates of the volume-averaged SFR at low
redshifts derived from various UV and H$\alpha$ measurements.  The
upper panel shows the `raw' estimates, whilst in the lower panel the UV
data (this work and that based on the rest-frame 2800 \AA\ flux density
of Lilly et al.~(1997)) has been corrected for a modest amount
of dust extinction following the discussion of Madau (1997). In both panels
the other estimates based on $H\alpha$ fluxes (Tresse \& Maddox 1997,
Gallego et al.~1995) are as discussed by Madau. Our dust-corrected
value is in good agreement with the Tresse \& Maddox determination, but
a factor of $~1.6$ higher than the Lilly et al.~estimate and a factor
of $~2.5$ higher than the local H$\alpha$ estimate of Gallego et al.
Clearly there is growing evidence that the local density of star
formation has been underestimated and this upward trends reduces the
strength of the evolution claimed. The point is reinforced to some
extent by the absence of any strong evolutionary trend within our own
data. Ultimately, with a larger sample and better quality spectra, it 
will be profitable to compare the results obtained from the UV continuum
and $H\alpha$ emission on a one-to-one basis.

\begin{figure}
\psfig{figure=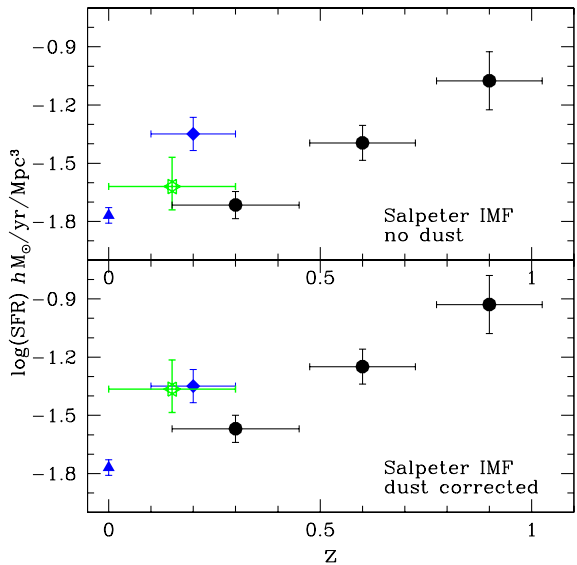,width=80mm}
\caption{The redshift dependence of the comoving volume-average star 
formation rate following Madau (1997). Filled circles represent UV 
(2800 \AA) broad band measurements based on CFRS (Lilly et al.~1996), the
filled triangle represents the Gallego et.~(1995) H$\alpha$ survey and 
the filled diamond the nebular line estimates based on CFRS (Tresse \& 
Maddox 1997). The star represents the value derived from the present 
UV (2000 \AA) survey. The conversion from a luminosity density assumes 
a Salpeter IMF with stellar masses ranging from 0.1 to 125 ${\rm M}_\odot$. 
In the lower panel, the observed rest-frame UV light for this work and
the Lilly et al.~data have been corrected upwards for dust extinction 
by a factor of 1.4 at 2800 \AA, and 1.8 at 2000 \AA\ respectively.
Other values are unchanged.}
\end{figure}

\section{Conclusions}

We present the first results of a ongoing spectroscopic survey of
galaxies selected in the rest-frame ultraviolet.  The source catalogue for this
survey was constructed from a flux-limited sample of stars, galaxies and QSOs
imaged at 2000 \AA\ with the FOCA balloon-borne camera (Milliard et al.~1992). 
Optical spectroscopy was conducted with the WIYN  and WHT telescopes for 
142 faint sources in Selected Area 57. 

The redshift distribution extends over $0<z<0.5$, with mean redshift 
$\bar z\sim 0.15$. A high fraction (46\%) of the sources shows intense 
nebular emission lines with $W_\lambda$[OII]3727\AA\ $>$ 15~\AA,
and about 10\% show ultraviolet-optical colors up to 2 magnitudes bluer 
than expected for normal Hubble sequence galaxies. 

Our data suggests that optically-selected samples used for deriving
$k$-corrections are deficient in star-forming galaxies. The additional
star-forming galaxies may contribute to the apparent excess of faint
blue sources observed in deep optical sky surveys and will reduce the
strength of the claimed evolution in blue luminosity density to
$z\simeq$1. 

The UV-optical colour distribution and UV number counts
cannot easily be reconciled via minor adjustments to the optical
luminosity distributions. In general, it seems the statistical
properties of galaxies in the UV cannot be accurately predicted from
limited UV data for optically-selected samples.

We have derived the rest-frame ultraviolet galaxy luminosity function
at the mean redshift of the sample. We find a faint end slope $\alpha
\sim - 1.6$ considerably steeper than that of the local optical
luminosity function.

Adopting a Salpeter IMF and a modest amount of dust reddening, we use
the integrated UV light detected to estimate the local volume averaged
star formation rate.  Our value is in good agreement with the recent
estimate of Tresse \& Maddox (1997), derived from H$\alpha$ line
measurements at similar redshifts, but a factor of 1.6 higher than the
low-redshift estimate of Lilly et al. (1996), based on optical data
colour shifted in the near UV.
 
\section*{ACKNOWLEDGEMENTS}

We thank the referee, Dr Gerhardt Meurer, for his valuable
comments and for pointing out an inconsistency in our original
manuscript. We thank Richard McMahon for invaluable assistance in
matching the UV data with APM catalogues and Di Harmer for
the charm and efficiency with which she helped us secure
the WIYN data. We acknowledge generous assistance from support
staff at La Palma and useful discussions with Tom Broadhurst,
Max Pettini and Laurence Tresse.

\begin{table*}
\caption{UV galaxies with measured redshift. OC is the number of optical 
counterparts within 10'' of the UV source if it was detected by FOCA 1500,
or within 20'' if it was detected by FOCA 1000 only. The position (equinox 
1950.0) is that of the observed counterpart. T is the galaxy type index 
inferred from the model UV-$b$ colour-redshift relation (as described in 
the text): T=1 to 8 refers to types E, S0, Sa, Sb, Sc, Sd, SB1 and SB2 
respectively. T=0 means no colour information is available for the galaxy.}
\begin{tabular}{@{}lrrrrrrrrrrl} 
sample & OC & RA ~~~~~~~DEC~~~ & UV~ & $B$~~ &  $z~~$ & OII  & H$\alpha$  & T   & $k_{UV}$ & $M_{UV}$ &Comments\\
\noalign{\hrule}
\input uvz_wiyn.tbl
\end{tabular}
\end{table*}

\begin{table*}
\begin{tabular}{@{}lrrrrrrrrrrl}
sample & OC & RA ~~~~~~~DEC~~~ & UV~ & $B$~~ &  $z~~$ & OII  & H$\alpha$  &T   & $k_{UV}$ & $M_{UV}$ &Comments \\
\noalign{\hrule}
\input uvz_wht.tbl

\end{tabular}
\end{table*}

\begin{table*}
\begin{tabular}{@{}lrrrrll}
sample & OC & RA ~~~~~~~DEC~~~ & UV~ & $B$~~ &  $z~~$ & Comments \\
\noalign{\hrule}
\input uvz_others.tbl
\end{tabular}
\end{table*}

\end{document}

%% file: uvz_wiyn.tbl
WIYN
 & 1 & 13:06:01.14 29:21:14.7 & 18.53 & 19.29 &  0.2407 &   0 &   0 &  4 & -0.19 &-20.69 & HK,Hd               \\
 & 1 & 13:04:31.58 28:48:55.0 & 17.68 & 10.72 &  0.0206 &   0 &   0 &  1 &  0.15 &-16.43 & HK,abs              \\
 & 1 & 13:05:01.02 29:03:43.2 & 17.75 & 18.54 &  0.1146 &   3 &   0 &  5 & -0.13 &-19.86 & OII, HK             \\
 & 1 & 13:06:27.79 29:09:09.3 & 16.77 & 14.77 &  0.0800 &   4 &   0 &  2 &  0.24 &-20.41 & OII,HK              \\
 & 1 & 13:06:08.41 29:16:38.7 & 18.19 & 18.69 &  0.0249 &  16 &   0 &  5 & -0.02 &-16.17 & OII, OIII           \\
 & 1 & 13:07:29.10 28:38:54.5 & 15.23 & 11.00 &  0.0231 &   0 &   0 &  1 &  0.16 &-19.15 & HK, abs             \\
 & 1 & 13:06:13.24 28:56:47.1 & 18.20 & 18.31 &  0.1222 &  15 &   0 &  4 & -0.13 &-19.56 & OII,HK,abs          \\
 & 1 & 13:05:13.62 28:44:45.9 & 18.15 & 18.65 &  0.0686 &  22 &   0 &  4 & -0.05 &-18.40 & OII,Hb,OIII         \\
 & 2 & 13:06:29.96 28:50:03.5 & 18.27 & 19.97 &  0.1849 &   6 &   0 &  6 & -0.20 &-20.34 & OII,HK,abs          \\
 & 1 & 13:06:26.18 28:50:23.5 & 17.22 & 17.30 &  0.0886 &  41 &   0 &  4 & -0.08 &-19.87 & OII,Hb,OIII         \\
 & 2 & 13:05:51.61 29:00:17.8 & 18.38 &  0.00 &  0.1053 &   0 &   0 &  0 & -0.11 &-19.06 & HK                  \\
 & 1 & 13:05:50.11 29:06:51.2 & 16.36 & 13.28 &  0.0544 &   0 &   0 &  2 &  0.18 &-19.91 & HK, abs             \\
 & 1 & 13:05:08.45 28:44:10.1 & 17.32 & 16.91 &  0.0685 &  11 &   0 &  3 & -0.03 &-19.25 & OII, abs            \\
 & 2 & 13:03:57.96 28:52:18.3 & 18.04 & 20.91 &  0.1865 &  47 &   0 &  8 & -0.29 &-20.50 & OII, OIII           \\
 & 1 & 13:05:13.02 28:41:19.4 & 18.51 & 19.82 &  0.0770 &  42 &   0 &  6 & -0.07 &-18.28 & OII, Hb, OIII       \\
 & 1 & 13:03:47.89 29:02:28.4 & 17.49 & 19.26 &  0.0836 &  42 &   0 &  7 & -0.11 &-19.44 & OII, OIII           \\
 & 1 & 13:07:03.27 28:59:17.8 & 17.99 & 19.03 &  0.0376 &  23 &   0 &  6 & -0.03 &-17.26 & OII, Hb             \\
 & 1 & 13:06:04.40 28:36:59.0 & 17.53 & 16.60 &  0.0237 &  25 &   0 &  3 &  0.00 &-16.74 & OII, Hb, OIII       \\
 & 1 & 13:07:40.56 28:56:33.1 & 18.21 & 19.46 &  0.1220 &  11 &   0 &  6 & -0.14 &-19.53 & OII, HK, abs        \\
 & 1 & 13:06:20.66 29:09:25.9 & 16.93 & 17.00 &  0.0612 &  35 &   0 &  4 & -0.04 &-19.38 & OII,Hb,OIII         \\
 & 1 & 13:06:14.98 29:10:25.8 & 16.59 & 17.55 &  0.0395 &  28 &   0 &  6 & -0.03 &-18.77 & OII,Hb,OIII         \\
 & 3 & 13:04:19.78 29:00:26.9 & 17.46 & 18.34 &  0.1127 &  18 &   0 &  5 & -0.12 &-20.12 & OII,Balmer,Hb,OIII  \\
 & 1 & 13:04:52.46 28:41:20.3 & 17.71 & 19.17 &  0.0698 &  17 &   0 &  6 & -0.06 &-18.87 & OII,Hb,OIII         \\
 & 2 & 13:04:45.12 28:41:37.3 & 18.04 & 18.59 &  0.0346 &  19 &   0 &  5 & -0.03 &-17.03 & OII                 \\
 & 1 & 13:03:55.90 28:44:11.1 & 17.40 & 18.92 &  0.2379 &  30 &   0 &  6 & -0.21 &-21.77 & OII,Balmer,Hb,OIII  \\
 & 1 & 13:05:45.10 29:18:35.4 & 16.59 & 17.14 &  0.0587 &  21 &   0 &  5 & -0.05 &-19.62 & OII,HK,Hb,OIII      \\
 & 2 & 13:07:05.25 29:11:28.7 & 16.82 & 17.77 &  0.0627 &  25 &   0 &  5 & -0.05 &-19.53 & OII,Hb,OIII         \\
 & 1 & 13:04:06.93 28:53:34.4 & 17.84 & 18.20 &  0.1866 &   5 &   0 &  4 & -0.18 &-20.81 & OII,HK,Hb,OIII      \\
 & 1 & 13:05:39.01 29:18:05.9 & 18.12 & 20.25 &  0.1443 &  32 &   0 &  7 & -0.22 &-19.92 & OII,OIII            \\
 & 1 & 13:05:54.39 28:57:38.7 & 16.57 & 19.25 &  0.0804 &  13 &   0 &  8 & -0.13 &-20.26 & OII,OIII            \\
 & 1 & 13:05:42.91 29:22:52.8 & 17.49 & 19.34 &  0.0525 &   6 &   0 &  7 & -0.06 &-18.47 & OII,OIII            \\
 & 1 & 13:06:03.73 28:50:27.4 & 16.41 & 16.29 &  0.0700 &  25 &   0 &  4 & -0.05 &-20.19 & OII,Balmer,Hb,OIII  \\
 & 1 & 13:06:49.06 28:43:19.2 & 17.91 & 18.59 &  0.0789 &  26 &   0 &  5 & -0.07 &-18.93 & OII, OIII           \\
 & 1 & 13:06:23.77 29:05:09.9 & 17.19 & 20.72 &  0.1384 &  31 &   0 &  8 & -0.23 &-20.74 & OII,Hb,OIII         \\
 & 1 & 13:07:26.48 29:14:32.2 & 16.79 & 19.89 &  0.1222 &  12 &   0 &  8 & -0.21 &-20.88 & OII                 \\
 & 1 & 13:04:48.14 28:41:26.5 & 17.95 & 18.92 &  0.0179 &  30 &   0 &  6 & -0.01 &-15.70 & OII, OIII           \\
 & 1 & 13:04:33.35 28:48:11.3 & 18.23 & 20.08 &  0.1589 &   0 &   0 &  6 & -0.19 &-20.05 & HK                  \\
 & 2 & 13:03:43.12 28:59:41.0 & 18.24 &  0.00 &  0.3712 &  11 &   0 &  0 & -0.40 &-21.76 & OII                 \\
 & 1 & 13:05:52.13 29:17:22.0 & 16.82 & 16.90 &  0.0206 &  25 &   0 &  4 & -0.01 &-17.13 & OII,Hb,OIII         \\
 & 1 & 13:05:41.78 28:58:41.9 & 16.51 & 15.64 &  0.0800 &   4 &   0 &  3 & -0.04 &-20.39 & OII,HK,abs          \\
 & 1 & 13:05:16.20 29:17:05.5 & 18.04 & 17.17 &  0.1235 &   4 &   0 &  3 & -0.10 &-19.77 & OII,HK,abs          \\
 & 1 & 13:06:32.30 28:51:22.4 & 18.27 & 20.00 &  0.0696 &  33 &   0 &  6 & -0.06 &-18.31 & OII,Hb,OIII         \\
 & 2 & 13:08:00.21 28:54:33.6 & 17.61 & 18.60 &  0.1223 &  11 &   0 &  5 & -0.14 &-20.13 & OII,HK              \\
 & 1 & 13:07:39.47 29:07:24.7 & 18.06 & 20.28 &  0.2348 &  14 &   0 &  7 & -0.28 &-21.01 & OII,HK              \\
 & 1 & 13:04:05.73 28:41:24.0 & 17.89 & 19.09 &  0.0695 &  23 &   0 &  6 & -0.06 &-18.68 & OII,HK              \\
 & 1 & 13:05:58.96 28:47:52.5 & 18.25 & 19.91 &  0.1993 &   8 &   0 &  6 & -0.21 &-20.52 & OII,HK              \\
 & 1 & 13:05:40.90 29:15:52.3 & 16.20 & 14.49 &  0.0242 &  23 &   0 &  3 & -0.01 &-18.11 & OII,Hb,OIII         \\
 & 1 & 13:04:57.97 29:15:45.5 & 17.81 & 19.25 &  0.1894 &  26 &   0 &  6 & -0.20 &-20.85 & OII,OIII            \\
 & 1 & 13:04:14.54 29:07:00.6 & 14.26 &  0.00 &  0.0160 &  24 &   0 &  0 & -0.01 &-19.14 & OII,Hb,OIII         \\
 & 1 & 13:06:34.50 28:59:59.0 & 17.60 & 12.18 &  0.0560 &   0 &   0 &  1 &  0.31 &-18.86 & HK, abs             \\
 & 1 & 13:06:28.46 28:53:45.1 & 16.16 & 13.63 &  0.0227 &  17 &   0 &  2 &  0.10 &-18.12 & OII,Balmer,OIII     \\
 & 1 & 13:06:24.43 28:53:25.3 & 16.66 & 17.33 &  0.1893 &  79 &   0 &  4 & -0.19 &-22.02 & OII,Hb,OIII         \\
 & 1 & 13:04:25.09 28:59:31.8 & 18.09 & 17.83 &  0.0565 &   5 &   0 &  3 & -0.02 &-18.06 & OII,HK              \\
 & 1 & 13:06:32.05 29:07:04.6 & 17.66 & 17.78 &  0.0794 &   4 &   0 &  4 & -0.06 &-19.20 & OII,HK, abs         \\
 & 1 & 13:04:14.22 28:51:15.7 & 17.26 & 16.82 &  0.1225 &   0 &   0 &  3 & -0.10 &-20.53 & HK,abs              \\

%% file: uvz_wht.tbl
WHT \#1
 & 2 & 13:03:58.95+28:52:21.8 & 18.02 & 21.02 &  0.2531 &  12 &   9 &  8 & -0.32 &-21.18 & OII                 \\
 & 1 & 13:04:24.07+29:06:57.9 & 17.33 & 17.62 &  0.0160 &   0 &   7 &  4 & -0.01 &-16.08 & Balmer, Ha          \\
 & 1 & 13:04:44.85+28:54:00.4 & 16.41 & 15.71 &  0.0393 &  25 &  26 &  3 & -0.01 &-18.96 & OII, OIII, Ha       \\
 & 2 & 13:05:30.34+29:03:16.3 & 17.70 & 20.53 &  0.2473 &  38 & 113 &  8 & -0.32 &-21.45 & OII, OIII           \\
 & 1 & 13:06:00.96+29:10:29.5 & 18.35 & 21.58 &  0.2702 &  49 &  -9 &  8 & -0.33 &-20.99 & OII, OIII           \\
 & 2 & 13:05:55.61+29:12:27.8 & 17.38 & 21.00 &  0.1959 &  41 &  32 &  8 & -0.30 &-21.26 & OII, OIII, Ha:      \\
 & 1 & 13:05:59.62+29:13:10.1 & 18.08 & 19.18 &  0.1757 &   8 &  13 &  5 & -0.20 &-20.42 & OII, Ha             \\
 & 1 & 13:06:01.98+29:15:06.0 & 17.42 & 19.11 &  0.0256 &  32 &  86 &  7 & -0.03 &-17.00 & 0II, Ha             \\
 & 1 & 13:05:32.99+29:16:56.8 & 18.17 & 19.62 &  0.1377 &   5 &   4 &  6 & -0.16 &-19.82 & OII, HK             \\
 & 2 & 13:05:22.71+29:21:19.4 & 17.91 & 21.50 &  0.1913 &   5 &  -9 &  8 & -0.29 &-20.69 & OII, Hb, OIII       \\
 & 1 & 13:04:29.55+29:22:03.1 & 17.86 & 20.16 &  0.0224 &   6 &  48 &  7 & -0.02 &-16.27 & OII, OIII:, Ha      \\
 & 2 & 13:05:38.86+29:32:23.6 & 18.22 & 21.68 &  0.0493 &   6 &   1 &  8 & -0.07 &-17.59 & OII, OIII, Ha:      \\
 & 2 & 13:04:53.90+29:27:21.0 & 17.82 & 21.67 &  0.5210 &  58 &   0 &  8 & -0.74 &-22.63 & OII, Hb, OIII       \\
 & 1 & 13:05:19.19+29:32:52.8 & 18.15 & 18.25 &  0.0848 &   5 &   6 &  4 & -0.07 &-18.85 & OII, HK, Ha         \\
 & 2 & 13:05:27.53+29:35:30.0 & 18.28 & 20.87 &  0.2525 &  15 &  12 &  8 & -0.32 &-20.91 & OII, Balmer, Ha     \\
 & 1 & 13:04:35.77+29:25:56.9 & 17.44 & 17.72 &  0.0607 &  27 &  29 &  4 & -0.04 &-18.85 & OII, OIII, Ha       \\
 & 1 & 13:05:10.07+29:35:40.6 & 18.04 & 17.89 &  0.1275 &   0 &   0 &  3 & -0.11 &-19.83 & Balmer only!        \\
 & 1 & 13:04:27.19+29:26:58.7 & 17.65 & 17.95 &  0.1382 &   3 &   7 &  4 & -0.15 &-20.36 & OII, HK, Ha         \\
 & 1 & 13:04:31.61+29:31:20.9 & 17.79 & 17.99 &  0.1379 &   9 &  35 &  4 & -0.15 &-20.21 & OII, HK, Ha         \\
 & 1 & 13:04:40.52+29:41:16.2 & 17.44 & 21.78 &  0.2677 &  16 &   0 &  8 & -0.33 &-21.88 & OII                 \\
 & 1 & 13:04:30.97+29:35:33.5 & 17.63 & 18.87 &  0.2256 &  13 &  25 &  5 & -0.21 &-21.42 & OII, HK             \\
 & 2 & 13:04:09.94+29:27:03.8 & 16.65 & 14.01 &  0.0312 &   0 &  13 &  2 &  0.12 &-18.35 & HK, Ha              \\
 & 1 & 13:03:54.86+29:33:35.6 & 18.20 & 20.69 &  0.2736 &  54 &  70 &  8 & -0.33 &-21.17 & OII, Hb, OIII, Ha   \\
 & 1 & 13:04:22.13+29:29:59.1 & 18.22 & 19.60 &  0.1823 &  11 &  12 &  6 & -0.20 &-20.36 & OII, HK, Ha         \\
 & 1 & 13:04:16.07+29:27:03.2 & 18.19 & 21.81 &  0.1850 &  32 &   0 &  8 & -0.29 &-20.34 & OII                 \\
 & 2 & 13:03:40.31+29:37:37.7 & 18.14 & 21.96 &  0.2858 &  14 &   0 &  8 & -0.35 &-21.31 & OII,OIII (poor ext.)  \\
 & 1 & 13:03:20.22+29:41:43.9 & 17.91 & 19.76 &  0.0894 &  50 &  49 &  7 & -0.12 &-19.16 & OII, OIII, Ha       \\
 & 1 & 13:03:36.41+29:32:45.2 & 18.13 & 11.33 &  0.0266 &   0 &   0 &  1 &  0.18 &-16.57 & Balmer only!        \\
 & 1 & 13:03:15.26+29:37:02.5 & 16.85 & 16.78 &  0.0238 &  11 &   6 &  4 & -0.01 &-17.42 & OII, HK             \\
 & 1 & 13:03:11.64+29:22:44.9 & 17.98 & 16.79 &  0.0819 &   0 &  13 &  3 & -0.04 &-18.97 & HK, Ha              \\
 & 1 & 13:03:08.58+29:21:40.8 & 17.98 & 21.63 &  0.0237 &   0 &  23 &  8 & -0.03 &-16.26 & Ha                  \\
 & 1 & 13:03:13.39+29:35:00.4 & 18.05 & 19.42 &  0.0897 &  28 &  33 &  6 & -0.09 &-19.06 & OII, OIII, Ha       \\
 & 1 & 13:03:23.02+29:31:13.0 & 18.23 & 20.86 &  0.2897 & 125 & 250 &  8 & -0.35 &-21.25 & OII, OIII, Ha       \\
 & 1 & 13:03:15.09+29:29:57.6 & 17.78 & 18.00 &  0.1390 &   2 &   4 &  4 & -0.15 &-20.24 & OII, HK, Ha         \\
 & 1 & 13:03:50.38+29:24:30.5 & 16.57 & 16.42 &  0.0235 &   0 &   4 &  4 & -0.01 &-17.67 & OIII, Ha            \\
 & 1 & 13:02:34.52+29:32:08.1 & 17.49 & 21.06 &  0.1682 &  -9 &  -9 &  8 & -0.27 &-20.84 & OII, Ha (poor ext.)   \\
 & 1 & 13:02:57.34+29:18:58.1 & 18.32 & 19.52 &  0.0176 &   0 &  28 &  6 & -0.01 &-15.29 & Ha, Hb, OIII, SII   \\
 & 2 & 13:02:52.59+29:16:59.9 & 18.49 & 21.97 &  0.0332 &   0 &   0 &  8 & -0.05 &-16.47 & HK, abs             \\
 & 1 & 13:02:32.35+29:12:59.0 & 18.09 & 19.86 &  0.2427 &  11 &  30 &  6 & -0.21 &-21.12 & OII, Balmer, Ha     \\
 & 2 & 13:02:53.91+29:08:50.9 & 18.39 & 19.88 &  0.3229 &  10 &  -9 &  5 & -0.28 &-21.41 & OII, OIII, Ha       \\
WHT \#2
 & 1 & 13:00:19.80 29:42:12.0 & 17.23 & 17.14 &  0.0900 &   8 &  27 &  4 & -0.08 &-19.90 & OII, H              \\
 & 1 & 12:59:17.64 29:38:59.6 & 16.12 & 15.71 &  0.0590 &  48 &  77 &  3 & -0.02 &-20.13 & OII, OIII, Ha       \\
 & 1 & 12:59:40.35 29:31:19.2 & 17.59 & 11.18 &  0.0250 &   0 &   0 &  1 &  0.17 &-16.97 & HK, abs             \\
 & 2 & 12:59:22.62 29:20:41.6 & 16.16 & 16.25 &  0.0620 &  21 &  20 &  4 & -0.04 &-20.18 & OII, Ha             \\
 & 1 & 12:59:33.16 29:19:06.8 & 17.44 & 19.47 &  0.0240 &  90 &  49 &  7 & -0.02 &-16.84 & OII, Ha             \\
 & 3 & 13:00:17.07 29:17:06.6 & 17.52 & 18.57 &  0.0380 &  17 &  18 &  6 & -0.03 &-17.75 & OII, OIII, Ha       \\
 & 1 & 13:02:06.03 29:11:51.5 & 17.32 & 17.07 &  0.0830 &   0 &  65 &  3 & -0.04 &-19.66 & OIII, Ha            \\
 & 1 & 13:02:03.66 29:18:39.3 & 18.06 & 17.80 &  0.0840 &   0 &  49 &  3 & -0.05 &-18.94 & HK, Ha              \\
 & 1 & 13:01:59.75 29:24:29.1 & 17.89 & 18.94 &  0.1890 &  12 &  21 &  5 & -0.20 &-20.77 & OII, HK, Ha         \\
 & 1 & 13:00:00.88 30:00:51.0 & 17.67 & 18.80 &  0.1570 &  10 &  13 &  5 & -0.18 &-20.59 & OII, Ha             \\

%% file: uvz_others.tbl
QS0'S
& 1 & 13:06:36.24 29:21:56.4 & 17.80 & 19.05 & 0.7460 & MgII   \\
& 1 & 13:07:21.82 28:43:32.7 & 16.41 & 18.24 & 0.7370 & MgII   	\\ 
& 1 & 13:04:41.44 29:17:32.8 & 17.69 & 19.50 & 1.5866 & CIV, CIII\\ 
& 1 & 13:04:23.17 28:39:55.6 & 17.17 & 19.12 & 0.9186 & MgII      \\  
& 1 & 13:04:12.79 29:35:29.7 & 18.30 & 20.41 & 1.0195 & CIII, Mg II\\ 
STARS
& 1 & 13:03:47.20 29:17:50.1 &  8.34 & -9.87 & $\star$    & $\star$  \\ 
& 1 & 13:00:08.49 29:23:14.1 & 17.50 & 14.50 & $\star$    & $\star$  \\
& 1 & 13:01:34.41 29:19:34.1 & 16.85 & 15.12 & $\star$    & $\star$  \\             	
& 1 & 13:07:33.80 28:49:27.4 & 15.48 & 13.33 & $\star$    & $\star$  \\            
& 1 & 13:07:56.30 29:02:17.9 & 18.47 & 20.04 & $\star$    & $\star$  \\              
& 2 & 13:05:13.95 29:04:36.3 & 15.29 & 14.07 & $\star$    & $\star$  \\              
& 2 & 13:07:52.45 28:48:54.9 & 17.56 & 21.43 & $\star$    & $\star$  \\              
& 1 & 13:04:58.22 29:11:27.1 & 18.45 & 13.93 & $\star$    & $\star$  \\           